\documentclass{llncs}

\usepackage{cmap}
\usepackage[utf8]{inputenc}
\usepackage{latexsym}
\usepackage[english]{babel}
\usepackage{graphicx,color}
\usepackage{hyperref}
\usepackage[all]{hypcap}
\usepackage{enumerate}
\usepackage{booktabs}
\usepackage{cite}
\usepackage{newunicodechar}
\usepackage{listings}

\newcommand{\inverse}[1]{\overline{#1}}

\newunicodechar{×}{\ensuremath\times}

\newunicodechar{→}{\ensuremath\rightarrow}

\newunicodechar{↓}{\ensuremath\leadsto}

\newunicodechar{¤}{\ensuremath\rightleftharpoons}

\newunicodechar{þ}{{\small \ensuremath{\,|\!\!>\,}}}

\author{Torben Ægidius Mogensen\inst{1}\orcidID{0000-0003-4862-9193}}

\institute{DIKU, University of Copenhagen\\ Universitetsparken 5, DK-2100 Copenhagen O,~~Denmark\\ \email{torbenm@di.ku.dk}} 

\title{The Reversible Intermediate Language L4,\\
  its Type System, and\\ its Bidirectional Small-Step Semantics}

\date{\today}

\begin{document}

\maketitle

\newcommand{\infer}[3]{\frac{\displaystyle{#1}}{\displaystyle{#2}}(\mathit{#3})}

\renewcommand{\qquad}{\quad~}

\begin{abstract}
\noindent
L4 is a generic, reversible, linearly-typed, low-level
language intended as an intermediate language for compiling reversible
programming languages to reversible machines.  L4 is intended to
replace and generalise RSSA, a reversible SSA form.  Inspiration is
drawn from other low-level reversible languages such as HSSA.

In this paper, we define L4 first informally, and then give a formal
type system and a bidirectional small-step semantics for L4: The rules
have direction indicators to determine forwards or backwards
execution, and the direction can be changed during execution.
\end{abstract}

\section{Introduction}
RSSA~\cite{Mogensen:PSI2015} is a reversible variant of the static
single-assignment form SSA~\cite{Alpern:1988,Rosen:1988}. RSSA has a
specific set of instructions corresponding to the instructions of an
earlier reversible intermediate language RIL~\cite{Mogensen:RC2015},
which limits its generality.  Furthermore, RSSA is untyped, so it
lacks the safety and specificity of statically typed languages.  HSSA
(Hybrid SSA)~\cite{Gail+Meyer:2024} generalisates RSSA both by not
keeping the instruction set fixed and by adding an irreversible
instruction that discards its input.  Additionally, HSSA uses a
notation that makes it explicit which variables are created,
eliminated, or just used by an instruction.  For L4, we use a notation
inspired by both RSSA and HSSA, but add a type system, and generalise
conditional entries and exits.  We thank Lukas Gail for discussions
about the design of HSSA and L4.  The name ``L4'' stands for ``Linear,
Low-Level Language''.

We first give an informal description of L4, then a formal type
system, and finally a bidirectional small-step semantics.

\section{Types}\label{types}

L4 uses both simple types and function types.  The set $T$ of simple
types is mostly unspecified, but must contain at least an unsigned
integer type.  Typically, $T$ will contain the type \texttt{0}, which
contain only the value 0, a Boolean type \texttt{b} that contains only
0 and 1, number types such as \texttt{u8} and \texttt{u64} (unsigned
8-bit and 64-bit numbers), and memory types such as \texttt{m64}
(byte-addressed memories using unsigned 64-bit addresses).  Note that
L4 allows a program to have multiple memories. $T^*$ is the set of
lists of simple types.  $\mathit{Values}(s)$ is the set of values in
the simple type $s$.

Function types are of the form $s → t ¤ u$, where $s,\,t,\,u \in
T^*$. A type $s → t ¤ u$ specifies that values/variables of type $s$
are used, but not eliminated, that values/variables of type $t$ are
eliminated and values/variables of type $u$ are created.  The type
notation is taken from Agni, a reversible array-programming
language~\cite{Mogensen:2021}.  We use such function types both for
predefined operations and for functions defined in a L4 program.

A type of form $t¤u$ is linear both in the forwards and backwards
direction.  The read-only parameters are similar to borrowed values in
Rust~\cite{matsakis:2014}.  This allows variables to be read (but not
modified) between their creation and elimination, but neither before
nor after.  Note that the inverse of creation is elimination, and the
inverse of read-only use is read-only use.

Examples of predefined operations are:

\begin{itemize}
\item \texttt{< : u64 u64 → b ¤ b}, which uses two integers,
  eliminates a Boolean and creates a Boolean which is the eliminated
  Boolean XOR'ed by the result of the comparison.
\item \texttt{+ : u64 → u64 ¤ u64}, which adds two integers, one
  of which is eliminated, and creates a new integer that is the sum.
\item \texttt{- : u64 → u64 ¤ u64}, which subtracts two integers, the
  first of which is eliminated, and creates a new integer that is the
  difference.
\item \texttt{memadd32: m64 u64 → u32 ¤ u32}, that fetches a 32-bit
  value from a 64-bit address in a memory and adds that to its 32-bit
  input, giving a 32-bit output.
\item \texttt{swap8 : u64 → m64 u8 ¤ m64 u8}, that uses a 64-bit
  address, eliminates a memory and an 8-bit value and creates a new
  memory and a new 8-bit value.  The new 8-bit value is the contents
  of the address in the eliminated memory, and the eliminated 8-bit
  value is the contents of the address in the created memory.  Note
  that this treatment of memory is without side effects: The old
  memory is explicitly eliminated and a new explicitly created.  The
  single-elimination property of variables ensures that the same
  memory can not be updated twice.
\end{itemize}

\noindent
We allow predefined operations to be overloaded.  For example,
\texttt{+} can be defined for multiple integer types. Arguments to
predefined operations are always explicitly typed, so this resolves
any ambiguity.

\section{Constants, Variables, and Arguments}

Constants are typed numbers, e.g., \texttt{0:b} or \texttt{42:u64}.
Variables are also explicitly typed, e.g., \texttt{x:u64} or
\texttt{c:b}.  We use $d$ to denote a typed constant or variable.
Arguments $a$ are lists of typed constants and variables, e.g.,
\texttt{1:u64 x:u64}.  The type of an argument list is a list of the
element types, e.g., \texttt{u64 u64}.

\section{Instructions}

An instruction that uses a predefined operation $i$ of type $s → t ¤
u$ has the form $a_t þ\, i~a_s þ a_u$, where $a_s$, $a_t$ and $a_u$ are
argument lists of type $s$, $t$, and $u$, respectively.  It eliminates
values of type $t$, uses values of type $s$ and creates values of type
$u$.  The symbol þ is inspired by the pipe operator \texttt{þ} in
functional languages such as F\#.  The notation is similar to that
used by HSSA, except we use left-to-right value flow where HSSA uses
right-to-left value flow.

The inverse of the instruction $a_t þ i~a_s þ a_u$ is $a_u þ\,
\inverse{i}~a_s þ a_t$.

An instruction can also be a call to a function $f: s → t ¤ u$ that is
defined in L4.  This has the form $a_t þ \texttt{call}~f~a_s þ a_u$ or
$a_t þ \texttt{uncall}~f~a_s þ a_u$.  \texttt{uncall} executes a
function backwards, so inputs become outputs and vice-versa.  The two
forms are mutually inverse.

When an instruction has constant outputs, this is an assertion that
the expected output is equal to the constant.

\section{Blocks, Entries and Exits}

A block consists of an entry, a (possibly empty) sequence of
instructions, and an exit.

An entry has the following form: $l_0\ldots l_n\, þ\, a$, where
$l_0\ldots l_n$ are labels and $a$ is an argument list.  This
represents a multi-way entry. The first argument is an integer control
variable that specifies which of the $n+1$ labels was used to enter
the block.  The remaining arguments are the arguments that were passed
in the jump to the block. It is common for an entry to have only one
label, in which case the first argument in $a$ is \texttt{0:0}, i.e.,
a constant value 0 of type \texttt{0}. This corresponds to an
unconditional entry.

An exit has the following form:

$a\, þ\, l_0\ldots l_n$, where $l_0\ldots l_n$ are labels and $a$ is
an argument list.  This represents a multi-way jump.  The first
argument is an integer control variable that specifies which of the
$n+1$ labels are jumped to.  The remaining arguments are passed to
this label.  Similar to entries, it is common to have only one label,
which repesents an unconditional jump.

This generalised form of entries and exits supports both conditional
and unconditional control flow, as well as multi-way jumps that act
similar to jump tables.  It reduces complexity of the language by
reducing the number of syntactic forms and semantic rules necessary.

A block is of the form $N~I_1 \ldots I_n~X$ where $N$ is an entry,
$I_1 \ldots I_n$ is a (possibly empty) list of instructions, and $X$
is an exit.

The labels \texttt{Begin}, \texttt{End}, and \texttt{ERROR} have
special meanings: \texttt{Begin} denotes the entry point of a function
and \texttt{End} its unique exit point. A jump to \texttt{ERROR} will
cause a run-time error.  Because of reversibility, \texttt{ERROR}
labels can also occur in entries.

\section{Function Definitions}

A function definition is of the form $\texttt{function}~f~a_s ~bs$ and
specifies a function $f$ of type $s → t ¤ u$.  $a_s$ specifies the
names and types of read-only variables used in $f$. This is akin to
the parameters of the parameterized relation in HSSA. $bs$ are the
blocks defining $f$.  $t$ is the list of types of the arguments to the
\texttt{Begin} label in $bs$, and $u$ is the list of types of the
arguments to the \texttt{End} label in $bs$.  Note that the order of
blocks in a function definition does not matter.

\section{Inverting Code}\label{inverting}

We can invert programs if any predefined operation $i$ of type $s → t
¤ u$ has an inverse $\inverse{i}$ of type $s → u ¤ t$.  Note that, in
the inverse, the arguments that are used, but not eliminated, are
still arguments that are used, but not eliminated, but the arguments
that in the original type are eliminated become results
that are created, and vice-versa.  Note that the inverse of \texttt{<}
is itself, the inverse of \texttt{+} is \texttt{-}, the inverse of
\texttt{swap8} is \texttt{swap8}, and the inverse of \texttt{memadd32}
is a function \texttt{memsub32} that subtracts a memory located value
from the eliminated argument.

Instructions, entries, exits, blocks, and function definitions can be
inverted using the following rules:

\begin{description}
\item[Instructions:]

The inverse of the instruction $a_t þ i~a_s þ a_u$ is $a_u þ
\inverse{i}~a_s þ a_t$.

The inverse of a call instruction $a_t þ \texttt{call}~f~a_s þ a_u$
is an uncall instruction $a_u þ \texttt{uncall}~f~a_s þ a_t$, and
vice-versa.

\item[Entries and exits:]

Entries are inverted to exits, and vice-versa. An entry of the form
$l_0\ldots l_n þ a$ is inverted to an exit of the form $a þ
\inverse{l_0}\ldots \inverse{l_n}$ and vice-versa, where
$\inverse{\texttt{Begin}} = \texttt{End}$, $\inverse{\texttt{End}} =
\texttt{Begin}$, and $\inverse{l} = l$ for all other labels.

\item[Blocks:]

A block of the form $N~I_1 \ldots I_n~X$ is inverted to
$\inverse{X}~\inverse{I_n}\ldots\inverse{I_1}~\inverse{N}$, where
$\inverse{N}$ is the inverse of the entry $N$, $\inverse{I_j}$
is the inverse of the instruction $I_j$, and $\inverse{X}$ is
the inverse of the exit $X$.
 
\item[Function definitions:]

A function definition of the form
$\texttt{function}\,f\,a_s\,b_1\cdots b_n$ is inverted to
$\texttt{function}\,f^{-1}\,a_s\,\inverse{b_1}\cdots \inverse{b_n}$,
where $\inverse{b}$ is the inverse of $b$.
\end{description}

\section{Syntax}\label{syntax}

A grammar for L4 is shown in Fig.~\ref{L4grammar}.

\begin{figure}
\[\begin{array}{lcl}
Program &→& Function^+\\\\
Function &→& \texttt{function}~F~A~~ B^+\\\\
B &→& N~I^*~X\\\\
N &→& L^+ þ A\\\\

X &→& A þ L^+\\\\

I &→& A þ P~A þ A \\
&|& A þ \texttt{call}~F~A þ A\\
&|& A þ \texttt{uncall}~F~A þ A\\\\

A &→& D^* \\
D &→& C:T\\
&|& Y:T
\end{array}\]

where $L$ is a label name, $Y$ is a variable name, $C$ is an integer
constant, $P$ is the name of a predefined operation, $T$ is a simple
type name, and $F$ is a function name. Label names are alphanumeric
starting with an upper-case letter, variable and type names are
alphanumeric starting with a lower-case letter, and names of
predefined operations can be non-empty sequences of non-blank ASCII
symbols, excluding \texttt{call} and \texttt{uncall}.  Function names
are usually written in lower case, but can (like operator names) be
sequences of arbitrary symbols.
\caption{Grammar for L4}\label{L4grammar}
\end{figure}

Additional static semantic rules are placed on a program.  Programs
that do no adhere to these rules are ill-formed:

No two function definitions can have the same name.

Labels and variables are local to function definitions, and within
every function definition,

\begin{itemize}
\item A variable $y$ must occur in exactly one creation context and
  exactly one elimination context.

  A creation context is $a_u$ in an instruction $a_t þ\, i~a_s þ a_u$,
  $a_t þ \texttt{call}~f~a_s þ a_u$, or $a_t þ \texttt{uncall}~f~a_s þ
  a_u$; $a$ in an entry  $l_0\ldots l_n þ a$.

  An elimination context is $a_t$ in an instruction $a_t þ i~a_s þ
  a_u$, $a_t þ \texttt{call}~f~a_s þ a_u$, or $a_t þ
  \texttt{uncall}~f~a_s þ a_u$; $a$ in an exit $a þ
  l_0\ldots l_n$.

  Note that the read-only arguments $a_s$ in a function definition are
  neither creation nor elimination contexts, so the function body must
  preserve these unchanged.
  
\item In an instruction $a_t þ i~a_s þ a_u$, $a_t þ
  \texttt{call}~f~a_s þ a_u$, or $a_t þ \texttt{uncall}~f~a_s
  þ a_u$, the sets of variables occurring in $a_s$, $a_t$, and
  $a_u$ are disjoint.
  
\item A label $l$ occurs in exactly one entry and one exit.
  Exceptions are the labels \texttt{Begin} and \texttt{End}, which
  occur in one entry and one exit, respectively, but not both, and
  \texttt{ERROR}, which can occur any number of times.  An attempt to
  jump to \texttt{ERROR} will cause a run-time error.
\end{itemize}

\noindent
Like in SSA form, every variable has a unique creation point, but
unlike SSA form, every variable also has a unique elimination point.
When a variable is used in an elimination context (such as when it is
an argument to a primitive instruction or a function call), it can not
be used in any other elimination context.  Between these, the variable
can be read, but not modified.  This, in effect, makes variables and
functions linear with read-only uses being similar to borrowed values
in Rust.

\section{Examples}

Using Haskell-like notation, a Fibonacci-pair function can be defined
as

\begin{lstlisting}
  fib2 n = fib3 (0, 1) n

  fib3 (a, b) n | n>0 = fib3 (b, a+b) (n-1)
  fib3 (a, b) 0 = (a, b)
\end{lstlisting}

\noindent
Fig.~\ref{fib2} shows L4 code for defining this.  Note that this uses
a loop instead of recursion.

\begin{figure}
  \begin{lstlisting}
function fib2

Begin |> 0:0 n:u64
0:0 0:u64 1:u64 n:u64 |> Fib2

Fib2 L1 |> c:b x:u64 y:u64 m:u64
c:b |> = 0:u64 x:u64 |> 1:b
0:b |> < 0:u64 m:u64 |> c1:b
c1:b x:u64 y:u64 m:u64 |> L3 L2

L2 |> 0:0 x1:u64 y1:u64 m1:u64
m1:u64 |> - 1:u64|> m2:u64
x1:u64 |> + y1:u64|> y2:u64
0:0 y1:u64 y2:u64 m2:u64 |> L1

L3 |> 0:0 a:u64 b:u64 0:u64
0:0 a:u64 b:u64 |> End

\end{lstlisting}
\caption{Fibonacci-pair function in L4}\label{fib2}
\end{figure}

To illustrate read-only parameters to a function, we define a
power-function equivalent to the Haskell function

\begin{lstlisting}
  power x p | x > 1 = power1 x p 1

  power1 x 0 y = y
  power1 x p y = power1 x (p-1) (x*y)
\end{lstlisting}

\noindent
Note that we need \lstinline{x > 1} for reversibility.  The equivalent
L4 code is shown in Fig.~\ref{power}.  Note that, instead of
explicitly verifying that \lstinline{x > 1}, the assertion that
\lstinline{y < 2} is false when the second block is entered through
\lstinline{Recur} will fail when \lstinline{x <= 1}.

\begin{figure}
\begin{lstlisting}
function power x:u64 p:u64

Begin |> 0:0
0:u64 |> + p:u64 |> p0:u64
0:0 p0:u64 1:u64 |> Power

Power Recur |> c:b p1:u64 y:u64
c:b |> < y:u64 2:u64 |> 1:b
0:b |> = 0:u64 p1:u64 |> c1:b
c1:b p1:u64 y:u64 |> Loop Exit

Exit |> 0:0 0:u64 xp:u64
0:0 xp:u64 |> End

Loop |> 0:0 p2:u64 y1:u64
p2:u64 |> - 1:u64 |> p3:u64
y1:u64 |> * x:u64 |> y2:u64
0:0 p3:u64 y2:u64 |> Recur
\end{lstlisting}
\caption{Power function in L4}\label{power}
\end{figure}

\begin{figure}
\begin{lstlisting}
function map size:u64

Begin |> 0:0 a:a64
0:0 0:u64 a:a64 |> Map

Loop Map |> c:b i:u64 a1:a64
c:b |> = 0:u64 i:u64 |> 0:b
0:b |> < i:u64 size:u64 |> c1:b
c1:b i:u64 a1:a64 |> Final Apply

Final |> 0:0 i1:u64 a2:a64
i1:u64 |> - size:u64 |> 0:u64
0:0 a2:a64 |> End

Apply |> 0:0 i2:u64 a3:a64
a3:a64 0:u64 |> swap i2:u64 |> a4:a64 v:u64
|> % v:u64 3:u64 |> c2:u64
c2:u64 v:u64 |> M0 M1 M2

M0 |> 0:0 v1:u64
v1:u64 |> call ff 3:u64 |> w:u64
0:0 w:u64 |> M01

M1 |> 0:0 v2:u64
v2:u64 |> call ff 4:u64 |> w1:u64
0:0 w1:u64 |> M11

M2 |> 0:0 v3:u64
v3:u64 |> call ff 7:u64 |> w2:u64
0:0 w2:u64 |> M21

M01 M11 M21 |> c3:u64 w3:u64
c3:u64  |> %1 w3:u64 3:u64 |>
a4:a64 w3:u64  |> swap i2:u64 |> a5:a64 0:u64
i2:u64 |> + 1:u64 |> i3:u64
0:0 i3:u64 a5:a64 |> Loop


function ff x:u64

Begin |> 0:0 v:u64
v:u64 |> * x:u64 |> w:u64
0:0 w:u64 |> End
\end{lstlisting}
\caption{Map function in L4}\label{map}
\end{figure}

Fig.~\ref{map} shows a larger example that uses both function calls
and three-way control.

It should be clear from the examples that L4 programs are neither easy
to read nor easy to code, but the intention is that L4 is used as an
intermediate form in compilers where this is not a major issue.

\section{A Type System for L4}

As mentioned in Sect.~\ref{types}, we have a set $T$ of simple types
and we have function types of the form $s → t ¤ u$, where $s,\,t,\,u
\in T^*$.

We have a set $P$ of predefined operator names and a set $\pi$ of
bindings of names in $P$ to function types.

We will also build and use a mapping $\phi$ from function names $F$ to
function types and for each function $f$ a mapping $\lambda$ from label
names $L$ to $(Y→T)×T^*$, where $Y$ is the set of variable names. An
element $\rho$ of $Y→T$ is an environment that binds variables in
scope to their types.  The second component of the pair is types of
the arguments to $L$.  For \texttt{Begin} and \texttt{End}, these are
the types of the arguments and results of $f$, respectively.

Type rules for L4 are shown in Fig.~\ref{L4types}.  The rules assume
that the syntactic constraints listed in Sect.~\ref{syntax} are
obeyed.

We use some helper rules to deal with environments: $t \vdash_A a :
\rho_1 ¤ \rho_2$ checks that the variables and constants in $a$ are
consistent with $t$, removes the variables from $\rho_1$ (where they
must be), giving $\rho_2$.  Equivalently, it adds the variables to
$\rho_2$ (where they can not already be), giving $\rho_1$.  So, by
reading the same rules left-to-tight or right-to-left, we can either
remove variables from or add variables to environments.  Note that
when checking read-only parameters, we ignore the environment where
these have been removed (as they shouldn't be).

Instructions map environments to environments, reversibly.  $\pi,\phi
\vdash_I i : \rho_1 ¤ \rho_2$ signifies that instruction $i$ can use
bindings in $(\rho_1 \cap \rho_2)$ while transforming $\rho_1$ to
$\rho_2$.  It can use predefined operators defined in $\pi$ and
functions defined in $\phi$.

Entries create new values (corresponding to those that were eliminated
in the jumps/calls to the entries).  Labels are in $\lambda$ bound to
environments for in-scope variables and to the types for their
arguments.  $\lambda,\,\phi \vdash_N n : \rho$ indicates that $\rho$
is the environment for variables right after the entry $n$.  The
environments for \texttt{begin} and \texttt{end} contain only the
read-only parameters to the function.  All labels occuring in the same
entry or exit have the same environment.

Exits are symmetric to entries, and so are their rules.
$\lambda,\,\phi \vdash_X x : \rho$ indicates that $\rho$ is the
environment for variables right before the exit $x$.

A block consists of an entry, a list of instructions and an exit.  We
use some helper rules for lists of instructions, but otherwise we
simply chain together an entry, instructions, and an exit.
$\pi,\,\phi,\lambda \vdash_B b$ indicates that the block $b$ is
type-correct with respect to the mappings $\pi,\,\phi,\lambda$.

A function definition is type correct if its parameters and results
agree with the types in $\phi$ and all its blocks are correct with
respect to $\lambda$, which must include correct bindings for
\texttt{Begin} and \texttt{End}.

We use some helper rules to check a list of blocks.

Note that $\lambda$ is not explicitly constructed.  The bindings in
$\lambda$ are given by the constraints in the rules for entries,
exits, and function definitions.  Any $\lambda$ that satisfies these
constraints will be acceptable, but in practice, we want the smallest
such $\lambda$.  This can be constructed by a work-set algorithm that
starts from the \texttt{Begin} block and adds bindings for labels as
they are encountered.  Alternatively, logic variables and unification
can be used as in Hindley-Milner type inference.

$\phi$ is more easily constructed from the declarations of functions
and their \texttt{Begin} and \texttt{End} blocks.

\begin{figure}%L4types
  
\[\resizebox{\hsize}{!}{$\begin{array}{c}
\infer{}{~\vdash_A ~ : \rho ¤ \rho}{EmptyT}\hspace{4em}
\infer{t\vdash_A v:\rho_1 ¤ \rho_3}%
{t_1~t\vdash_A c{:}t_1~a : \rho_1 ¤ \rho_3}{ConstantT}\\[3ex]
      
\infer{\rho_1 = \rho_2 \cup \{y{:}t_1\} \qquad y \notin dom(\rho_2)
  \qquad t\vdash_A a : \rho_2 ¤ \rho_3}%
{t_1~t\vdash_A y{:}t_1~a : \rho_1 ¤ \rho_3}{VariableT}\\[6ex]
      
\infer{t \vdash_A a_t : \rho_1 ¤ \rho_3
\qquad s \vdash_A a_s : \rho_3  ¤ \rho'
\qquad u \vdash_A a_u : \rho_2 ¤ \rho_3
\qquad (p : s → t ¤ u) \in \pi}%
{\pi,\phi \vdash_I a_t þ p~a_s þ a_u : \rho_1 ¤ \rho_2}{PredefT}\\[3ex]

\infer{t \vdash_A a_t : \rho_1 ¤ \rho_3
\qquad s \vdash_A a_s : \rho_3 ¤ \rho'
\qquad u \vdash_A a_u : \rho_2 ¤ \rho_3
\qquad \phi\,f = s → t ¤ u}%
{\pi,\phi \vdash_I a_t þ \texttt{call}~f~a_s þ a_u : \rho_1 ¤ \rho_2}{CallT}\\[3ex]
\infer{\phi\,f = s → u ¤ t
\qquad t \vdash_A a_t : \rho_1 ¤ \rho_3
\qquad s \vdash_A a_s : \rho_3 ¤ \rho'
\qquad u \vdash_A a_u : \rho_2 ¤ \rho_3}%
{\pi,\phi \vdash_I a_t þ \texttt{uncall}~f~a_s þ a_u : \rho_1 ¤ \rho_2}{UncallT}\\[6ex]

\infer{}{\pi,\,\phi \vdash_J ~ : \rho ¤ \rho}{EmptyInstructionsT}\\[3ex]

\infer{\pi,\,\phi \vdash_I i : \rho_1 ¤ \rho_2
\qquad \pi,\,\phi \vdash_J j : \rho_2 ¤ \rho_3}%
{\pi,\,\phi \vdash_J i~j : \rho_1 ¤ \rho_3}{NonEmptyInstructionsT}\\[5ex]

\infer{\forall i\in[0,n], l_i\neq \texttt{ERROR}:\lambda\,l_i = (\rho,\,t)
\qquad s\,t \vdash_A a : \rho_1 ¤ \rho
\qquad [0,\, n] \subseteq \mathit{Values}(s)}%
{\lambda \vdash_N l_0\dots l_1 \rhd a : \rho_1}{EntryT}\\[6ex]

\infer{\forall i\in[0,n], l_i\neq \texttt{ERROR}:\lambda\,l_i = (\rho,\,t)
\qquad s\,t \vdash_A a : \rho_1 ¤ \rho
\qquad [0,\, n] \subseteq \mathit{Values}(s)}%
{\lambda \vdash_N a \rhd l_0\dots l_1: \rho_1}{ExitT}\\[6ex]

\infer{\lambda\vdash_N n : \rho_1
\qquad \pi,\,\phi \vdash_J j : \rho_1 ¤ \rho_2
\qquad \lambda \vdash_X x : \rho_2}%
{\pi,\,\phi,\lambda \vdash_B n~ j~ x}{BlockT}\\[6ex]

\infer{}{\pi,\,\phi,\lambda \vdash_{Bs} ~ }{EmptyBlocksT}\hspace{3em}

\infer{\pi,\,\phi,\lambda \vdash_B b
\qquad \pi,\,\phi,\lambda \vdash_{Bs}~bs}%
{\pi,\,\phi,\lambda \vdash_{Bs} b~bs}{NonEmptyBlocksT}\\[6ex]

\infer{{\begin{array}{c}\phi\,f = s → t ¤ u
\qquad s \vdash_A a_s : \rho_s ¤ \emptyset\\
\lambda\, \texttt{Begin} = (\rho_s,\,t) \qquad \lambda\,\texttt{End} = (\rho_s,\,u)
\qquad \pi,\,\phi,\lambda \vdash_{Bs} bs
\end{array}}}%
      {\pi,\,\phi \vdash_F \texttt{function}~f~a_s~~bs}{FunctionT}\\[6ex]

\infer{\vdash_{\inverse{G}} \inverse{g} : \phi
\qquad \pi,\,\phi \vdash_{\inverse{F}} \inverse{g}}%
{\pi,\,\vdash_{Program} \inverse{g}}{Program}\\[6ex]

\infer{\begin{array}{c}
 (l_0\ldots l_i \rhd a_t~j_1~x) \in bs \qquad \texttt{Begin} \in \{l_0,\,\ldots,\, l_i\}\\
\qquad (n~j_2~a_u\rhd l'_0 \ldots l'_k) \in bs \qquad \texttt{End}\in \{l'_0,\,\ldots,\, l'_k\}\\
\qquad s \vdash_A a_s : \rho_s ¤ \emptyset
\qquad t \vdash_A a_t : \rho_t ¤ \emptyset
\qquad u \vdash_A a_u : \rho_u ¤ \emptyset
\end{array}}%
{\vdash_G \texttt{function}~f~a_s~bs : \{f : t → s → u\}}{FunctionG}\\[6ex]

\infer{}{\vdash_{\inverse{G}} ~ : \emptyset}{EmptyG}\hspace{4em}

\infer{\vdash_G g : \phi_1\qquad \vdash_{\inverse{G}}\, \inverse{g} : \phi_2}%
{\vdash_{\inverse{G}} g ~ \inverse{g} : \phi_1\cup\phi_2}{NonEmptyG}\\[3ex]

\infer{}{\vdash_{\inverse{F}} ~ }{EmptyF}\hspace{7em}

\infer{\vdash_F g \qquad \vdash_{\inverse{F}}\, \inverse{g} }%
{\vdash_{\inverse{F}} g ~ \inverse{g}}{NonEmptyF}\\[3ex]

\end{array}$}\]
\caption{Type rules for L4}\label{L4types} 
\end{figure}

%%%%%%%%%%%%%%%%%%%%%%%%%%%%%%%%%%%%%%%%%%%%%%%%%%

\section{A Bidirectional Small-Step Semantics for L4}

In the following, we assume that programs are syntactially correct and
well typed.

A small-step semantics maps states to states with bounded changes.  A
state during execution of L4 consists of the following components:

\begin{enumerate}[1.]
\item An environment $\rho$ that maps currently live local variables
  to values.
\item A current program point $(n,\,i_0,\,i_1,\,x)$, which consists of
  the most recently entered block entry $n$, a list $i_0$ of the
  instructions already executed in the block, a list $i_1$ of
  instructions not yet executed in the block, and the exit $x$ of the
  block.
\item A direction indicator: $\gg$ for forwards execution and $\ll$ for
  backwards execution.
\item A current function name $f$.
\item A call stack $\sigma$, where each element is a tuple of an
  environment, a program point, a direction indicator, and a function
  name.
\end{enumerate}

\noindent
Additionally, we use mappings $\pi$, $\phi$, $\gamma_n$, and
$\gamma_x$, where $\pi$ maps the name $p$ of a predefined operation to
a function of type $s→t→u$\footnote{We assume overloading has been
  resolved at this time.}, $\phi$ maps a function name to the list of
read-only arguments to the function, $\gamma_n$ maps a function name
$f$ and a label name $l$ to the block that in $f$ has $l$ as an entry
label, and $\gamma_x$ maps a function name $f$ and a label name $l$ to
the block that in $f$ has $l$ as an exit label.  These mappings do not
change during execution. Note that $\overline{p}$ is the inverse to
the predfined operation $p$.  We show rules for both forwards
execution (indicated by $\gg$) and backwards execution (indicated by
$\ll$).  We need the rules to coexist to handle \texttt{uncall}.

\begin{figure}
\[\begin{array}{c}
\infer{}{\vdash^A : \rho ¤ \rho}{Empty}\qquad\qquad

\infer{w \vdash^A a : \rho_1 ¤ \rho_2\qquad V(c,t) = v
}{
  v\,w \vdash^A c{:}t\,\,a : \rho_1 ¤ \rho_2}{Constant}\\[4ex]

\infer{w \vdash^A a : \rho_1 ¤ \rho_2\qquad \rho_2 = \rho_3 \cup \{y\mapsto v\}\qquad
  y \notin dom(\rho_3)
}{
  v\,w \vdash^A y{:}t\,\,a : \rho_1 ¤ \rho_3}{Variable}\\[6ex]

\infer{\begin{array}{c}
  v_t \vdash^A a_t : \rho_0 ¤ \rho_1\qquad
  v_s \vdash^A a_s : \rho_1 ¤ \rho_2\\
  \pi(p)(v_s)(v_t) = v_u\qquad
  v_u \vdash^A a_u : \rho_3 ¤ \rho_1
  \end{array}
}{
  \pi,\,\phi,\,\gamma_n,\,\gamma_x \vdash\begin{array}{c}
  (\rho_0,\,(n,i_0,(a_tþp\,a_sþa_u)\,i_1,\,x),\,\gg,\,f,\,\sigma)\\ ↓\\
  (\rho_3,\,(n,i_0\,(a_tþp\,a_sþa_u),\,i_1\,x),\,\gg,\,f,\,\sigma)
  \end{array}
}{PredefF}\\[12ex]

\infer{\begin{array}{c}
  v_u \vdash^A a_u : \rho_0 ¤ \rho_1\qquad
  v_s \vdash^A a_s : \rho_1 ¤ \rho_2\\
  \pi(\overline{p})(v_s)(v_u) = v_t\qquad
  v_t \vdash^A a_t : \rho_3 ¤ \rho_1
  \end{array}
}{
  \pi,\,\phi,\,\gamma_n,\,\gamma_x \vdash\begin{array}{c}
  (\rho_0,\,(n,i_0\,(a_tþp\,a_sþa_u),\,i_1\,x),\,\ll,\,f,\,\sigma)\\ ↓\\
  (\rho_3,\,(n,i_0,(a_tþp\,a_sþa_u)\,i_1,\,x),\,\ll,\,f,\,\sigma)
  \end{array}
}{PredefB}\\[12ex]

\infer{\begin{array}{c}
    c\,v \vdash^A a : \rho_0 ¤ \rho_1\qquad 0\leq c \leq n\qquad
    l_c \notin \{\texttt{End},\,\texttt{ERROR}\}\\
    \gamma_n\,(f,\,l_c) = (l'_0\ldots l'_m þ a',\,i_1,\,x) \qquad
    l'_k = l_c  \qquad
    k\,v \vdash^A a' : \rho_2 ¤ \rho_1
  \end{array}
}{
  \pi,\,\phi,\,\gamma_n,\,\gamma_x \vdash\begin{array}{c}
  (\rho_0,\,(n,i_0,~,\,a þ l_0\ldots l_n),\,\gg,\,f,\,\sigma)\\ ↓\\
  (\rho_2,\,(l'_0\ldots l'_m þ a',~,\,i_1\,x),\,\gg,\,f,\,\sigma)
  \end{array}
}{JumpF}\\[12ex]

\infer{\begin{array}{c}
    c\,v \vdash^A a : \rho_0 ¤ \rho_1\qquad 0\leq c \leq n\qquad
    l_c \notin \{\texttt{Begin},\,\texttt{ERROR}\}\\
    \gamma_x\,(f,\,l_c) = (n,i_0,\,a' þ l'_0\ldots l'_m) \qquad
    l'_k = l_c  \qquad
    k\,v \vdash^A a' : \rho_2 ¤ \rho_1
  \end{array}
}{
  \pi,\,\phi,\,\gamma_n,\,\gamma_x \vdash\begin{array}{c}
  (\rho_0,\,(l_0\ldots l_n þ a,~,\,i_1\,x),\,\ll,\,f,\,\sigma)\\ ↓\\
  (\rho_2,\,(n,i_0,~,\,a' þ l'_0\ldots l'_m),\,\ll,\,f,\,\sigma)
  \end{array}
}{JumpB}\\[12ex]

\infer{\begin{array}{c}
    \gamma_n\,(f_0,\,\texttt{Begin}) = (l_0\ldots l_nþa,\,i_0,\,x)\qquad
    \phi\,f_0 = a_s\\
    w \vdash^A a_s : \rho_1 ¤ []\qquad
    c\,v \vdash^A a : \rho_0 ¤ \rho_1\qquad 0\leq c \leq n\qquad
    l_c = \texttt{Begin}
  \end{array}
}{
  \pi,\,\phi,\,\gamma_n,\,\gamma_x \vdash
  \texttt{start}(w,\,v) ↓
  (\rho_0,\,(l_0\ldots l_nþa,~,\,i_0,\,x),\,\gg,\,f_0,\,[])
}{StartF}\\[6ex]

\infer{\begin{array}{c}
    \gamma_x\,(f_0,\,\texttt{End}) = (n,\,i_0,\,a þ l_0\ldots l_n)\qquad
    \phi\,f_0 = a_s\\
    w \vdash^A a_s : \rho_1 ¤ []\qquad
    c\,v \vdash^A a : \rho_0 ¤ \rho_1\qquad 0\leq c \leq n\qquad
    l_c = \texttt{End}
  \end{array}
}{
  \pi,\,\phi,\,\gamma_n,\,\gamma_x \vdash
  \texttt{stop}(w,\,v) ↓
  (\rho_0,\,(n,\,i_0,~,\,a þ l_0\ldots l_n),\,\ll,\,f_0,\,[])
}{StartB}\\[6ex]

\infer{\begin{array}{c}
    c\,v \vdash^A a : \rho_0 ¤ \rho_1\qquad 0\leq c \leq n\qquad
    l_c = \texttt{End}\qquad
    \phi\,f_0 = a_s\qquad
    w \vdash^A a_s : \rho_1 ¤ []
  \end{array}
}{
  \pi,\,\phi,\,\gamma_n,\,\gamma_x \vdash
  (\rho_0,\,(n,i_0,~,\,a þ l_0\ldots l_n),\,\gg,\,f_0,\,[]) ↓
  \texttt{stop}(w,\,v)
}{StopF}\\[6ex]

\infer{\begin{array}{c}
    c\,v \vdash^A a : \rho_0 ¤ \rho_1\qquad 0\leq c \leq n\qquad
    l_c = \texttt{Begin}\qquad
    \phi\,f_0 = a_s\qquad
    w \vdash^A a_s : \rho_1 ¤ []
  \end{array}
}{
  \pi,\,\phi,\,\gamma_n,\,\gamma_x \vdash
  (\rho_0,\,(l_0\ldots l_nþa,~,\,i_0,\,x),\,\ll,\,f_0,\,[]) ↓
  \texttt{start}(w,\,v)
}{StopB}

\end{array}\]

\caption{A bidirectional small-step semantics for L4 (part 1)}
\label{bidirectional1}
\end{figure}

\begin{figure}
\[\begin{array}{c}

\infer{\begin{array}{c}
    v \vdash^A a'_t : \rho_0 ¤ \rho_1\qquad
    w \vdash^A a'_s : \rho_1 ¤ \rho_2\\
    \phi\,f = a_s\qquad \gamma_n\,(f,\,\texttt{Begin}) = (l_0\ldots l_nþa,\,i_0,\,x)\\
    
    w \vdash^A a_s : \rho_3 ¤ []\qquad
    l_c = \texttt{Begin}\qquad
    c\,v \vdash^A a : \rho_4 ¤ \rho_3\qquad\\
    p = (\rho_1,\,(n',i'_0\,\,a'_tþ\texttt{call } f\,a'_sþa'_u,\,i'_1,x'),\,\gg,\,g)
  \end{array}
}{
  \pi,\,\phi,\,\gamma_n,\,\gamma_x \vdash\begin{array}{c}
  (\rho_0,\, (n',i'_0,\,\,a'_tþ\texttt{call } f\,a'_sþa'_u\,\,i'_1,\,x'),\,\gg,\,g,\,\sigma)\\ ↓\\
  (\rho_4,\,(l_0\ldots l_nþa,~,i_0,\,x),\,\gg,\,f,\,p::\sigma)
  \end{array}
}{CallF}\\[12ex]

\infer{\begin{array}{c}
    v \vdash^A a'_u : \rho_0 ¤ \rho_1\qquad
    w \vdash^A a'_s : \rho_1 ¤ \rho_2\\
    \phi\,f = a_s\qquad \gamma_x\,(f,\,\texttt{End}) = (n,\,i_0,\,a þ l_0\ldots l_n)\\
    
    w \vdash^A a_s : \rho_3 ¤ []\qquad
    l_c = \texttt{End}\qquad
    c\,v \vdash^A a : \rho_4 ¤ \rho_3\qquad\\
    p = (\rho_1,\,(n',i'_0\,\,a'_tþ\texttt{call } f\,a'_s þ a'_u,\,\,i'_1,x'),\,\ll,\,g)
  \end{array}
}{
  \pi,\,\phi,\,\gamma_n,\,\gamma_x \vdash\begin{array}{c}
  (\rho_0,\, (n',i'_0\,\,a'_tþ\texttt{call } f\,a'_sþa'_u,\,\,i'_1,\,x'),\,\ll,\,g,\,\sigma)\\ ↓\\
  (\rho_4,\,(n,~,\,i_0,\,a þ l_0\ldots l_n),\,\ll,\,f,\,p::\sigma)
  \end{array}
}{CallB}\\[12ex]

\infer{\begin{array}{c}
    v \vdash^A a'_t : \rho_0 ¤ \rho_1\qquad
    w \vdash^A a'_s : \rho_1 ¤ \rho_2\\
    \phi\,f = a_s\qquad \gamma_x\,(f,\,\texttt{End}) = (n,\,i_0,\,a þ l_0\ldots l_n)\\
    
    w \vdash^A a_s : \rho_3 ¤ []\qquad
    l_c = \texttt{End}\qquad
    c\,v \vdash^A a : \rho_4 ¤ \rho_3\qquad\\
    p = (\rho_1,\,(n',i'_0,\,\,a'_tþ\texttt{uncall } f\,a'_s þ a'_u\,\,i'_1,x'),\,\gg,\,g)
  \end{array}
}{
  \pi,\,\phi,\,\gamma_n,\,\gamma_x \vdash\begin{array}{c}
  (\rho_0,\, (n',i'_0,\,\,a'_tþ\texttt{uncall } f\,a'_sþa'_u\,\,i'_1,\,x'),\,\gg,\,g,\,\sigma)\\ ↓\\
  (\rho_4,\,(n,~,\,i_0,\,a þ l_0\ldots l_n),\,\ll,\,f,\,p::\sigma)
  \end{array}
}{UncallF}\\[12ex]

\infer{\begin{array}{c}
    v \vdash^A a'_u : \rho_0 ¤ \rho_1\qquad
    w \vdash^A a'_s : \rho_1 ¤ \rho_2\\
    \phi\,f = a_s\qquad \gamma_n\,(f,\,\texttt{Begin}) = (l_0\ldots l_nþa,\,i_0,\,x)\\
    
    w \vdash^A a_s : \rho_3 ¤ []\qquad
    l_c = \texttt{Begin}\qquad
    c\,v \vdash^A a : \rho_4 ¤ \rho_3\qquad\\
    p = (\rho_2,\,(n',i'_0\,\,a'_tþ\texttt{uncall } f\,a'_sþa'_u,\,i'_1,x'),\,\ll,\,g)
  \end{array}
}{
  \pi,\,\phi,\,\gamma_n,\,\gamma_x \vdash\begin{array}{c}
  (\rho_0,\, (n',i'_0,\,\,a'_tþ\texttt{uncall } f\,a'_sþa'_u\,\,i'_1,\,x'),\,\ll,\,g,\,\sigma)\\ ↓\\
  (\rho_4,\,(l_0\ldots l_nþa,\,,i_0,~,\,x),\,\gg,\,f,\,p::\sigma)
  \end{array}
}{UncallB}\\[12ex]

\infer{\begin{array}{c}
    c\,v \vdash^A a : \rho_0 ¤ \rho_1\qquad
    l_c = \texttt{End}\qquad
    \phi\,f = a_s\qquad
    w \vdash^A a_s : \rho_1 ¤ []\\
    p = (\rho_2,\,(n',i'_0\,\,a'_tþ\texttt{call } f\,a'_sþa'_u,\,i'_1,x'),\,\gg,\,g)\\
    w \vdash^A a'_s : \rho_2 ¤ \rho_3\qquad v \vdash^A a'_u : \rho_4 ¤ \rho_2
  \end{array}
}{
  \pi,\,\phi,\,\gamma_n,\,\gamma_x \vdash\begin{array}{c}
  (\rho_0,\,(n,i_0,~,\,a þ l_0\ldots l_n),\,\gg,\,f,\,p::\sigma)\\ ↓\\
  (\rho_4,\, (n',i'_0\,\,a'_tþ\texttt{call } f\,a'_sþa'_u,\,i'_1,x'),\,\gg,\,g,\,\sigma)
  \end{array}
}{ReturnCallF}\\[12ex]

\infer{\begin{array}{c}
    c\,v \vdash^A a : \rho_0 ¤ \rho_1\qquad
    l_c = \texttt{Begin}\qquad
    \phi\,f = a_s\qquad
    w \vdash^A a_s : \rho_1 ¤ []\\
    p = (\rho_2,\,(n',i'_0\,\,a'_tþ\texttt{call } f\,a'_sþa'_u,\,i'_1,x'),\,\ll,\,g)\\
    w \vdash^A a'_s : \rho_2 ¤ \rho_3\qquad v \vdash^A a'_t : \rho_4 ¤ \rho_2
  \end{array}
}{
  \pi,\,\phi,\,\gamma_n,\,\gamma_x \vdash\begin{array}{c}
  (\rho_0,\,(l_0\ldots l_n þ a,~,i_1,\,x),\,\ll,\,f,\,p::\sigma)\\ ↓\\
  (\rho_4,\, (n',i'_0,\,\,a'_tþ\texttt{call } f\,a'_sþa'_u\,\,i'_1,x'),\,\ll,\,g,\,\sigma)
  \end{array}
}{ReturnCallB}
\end{array}\]

\caption{A bidirectional small-step semantics for L4 (part 2)}
\label{bidirectional2}
\end{figure}

\begin{figure}
\[\begin{array}{c}

\infer{\begin{array}{c}
    c\,v \vdash^A a : \rho_0 ¤ \rho_1\qquad
    l_c = \texttt{Begin}\qquad
    \phi\,f = a_s\qquad
    w \vdash^A a_s : \rho_1 ¤ []\\
    p = (\rho_2,\,(n',i'_0\,\,a'_tþ\texttt{uncall } f\,a'_sþa'_u,\,i'_1,x'),\,\gg,\,g)\\
    w \vdash^A a'_s : \rho_2 ¤ \rho_3\qquad v \vdash^A a'_u : \rho_4 ¤ \rho_2
  \end{array}
}{
  \pi,\,\phi,\,\gamma_n,\,\gamma_x \vdash\begin{array}{c}
  (\rho_0,\,(l_0\ldots l_n þ a,~,i_0,\,x)\,\ll,\,f,\,p::\sigma)\\ ↓\\
  (\rho_4,\, (n',i'_0\,\,a'_tþ\texttt{uncall } f\,a'_sþa'_u,\,i'_1,x'),\,\gg,\,g,\,\sigma)
  \end{array}
}{ReturnUncallF}\\[12ex]

\infer{\begin{array}{c}
    c\,v \vdash^A a : \rho_0 ¤ \rho_1\qquad
    l_c = \texttt{End}\qquad
    \phi\,f = a_s\qquad
    w \vdash^A a_s : \rho_1 ¤ []\\
    p = (\rho_2,\,(n',i'_0\,\,a'_tþ\texttt{uncall } f\,a'_sþa'_u,\,i'_1,x'),\,\ll,\,g)\\
    w \vdash^A a'_s : \rho_2 ¤ \rho_3\qquad v \vdash^A a'_u : \rho_2 ¤ \rho_4
  \end{array}
}{
  \pi,\,\phi,\,\gamma_n,\,\gamma_x \vdash\begin{array}{c}
  (\rho_0,\,(n,i_0,~,\,a þ l_0\ldots l_n),\,\gg,\,f,\,p::\sigma)\\ ↓\\
  (\rho_4,\, (n',i'_0,\,\,a'_tþ\texttt{uncall } f\,a'_sþa'_u\,\,i'_1,x'),\,\ll,\,g,\,\sigma)
  \end{array}
}{ReturnUncallB}

\end{array}\]

\caption{A bidirectional small-step semantics for L4 (part 3)}
\label{bidirectional3}
\end{figure}

In Fig.~\ref{bidirectional1}, we start by rules for using declarations
to remove or add values and variables.  As in the type rules,
$\vdash^A$ can be used both to remove variables from or add variables
to an environment.  It really just describes how an argument list
defines a relation between two environments and a list of values.

Note the $V(c,t)=v$ assertion in the rule $\mathit{(Constant)}$.  This
can fail if a value does not match a constant.

We then show rules for starting and stopping programs for forwards and
backwards execution.  $f_0$ is the first function in the program.
Note that reaching \texttt{End} (or \texttt{Begin} when running
backwards) with an empty call stack stops execution.  Rules for
non-empty call stacks are shown in Figs.~\ref{bidirectional2}
and~\ref{bidirectional3}.

the rule $\mathit{PredefF}$ describes the
effect of executing a predefined operation.  We first use the helper
rules $\vdash^A$ to find the values of the eliminated and read-only
arguments, while removing the eliminated variables from the
environment.  We then apply $\pi(p)$ to these arguments to get the
created values, and then bind the created values to
the created variables.  As a net effect, we update the environment and
move the current instruction from the list of not-yet executed
instructions to the already-executed instructions.  $\mathit{PredefB}$
describes backwards execution of a predefined operation.  Note the
symmetry between the two rules.

The rule $\mathit{JumpF}$ handles jumps between blocks. If the list of
not-yet-executed instructions is empty, we evaluate the arguments $a$
to the exit.  The first value, $c$, is the control argument, which is
used to select the label, $l_c$ for the jump, which must be different
from \texttt{End} and \texttt{ERROR}.  We use $\gamma_n$ to find the
destination block for $l_c$, find the label $l'_k = l_c$ in the entry
to this block.  Finally, we update the environment by binding the
control value $k$ and the non-control parameter values to the
parameter variables of the new block and change the state to begin
executing the new block.  $\mathit{JumpB}$ handles backwards jumps and
is symmetric to $\mathit{JumpF}$.  A jump can fail if the value of the
control parameter is not between 0 and the number of labels or if the
selected label is \texttt{ERROR}.

Jumps to \texttt{Begin} or \texttt{End} only happen at function calls
or returns or when a program starts or stops, so they are handled in
other rules.  Rules for starts and stops are shown in
Fig.~\ref{bidirectional1}. To run a program forwards, the initial state
is $\texttt{start}(w,\,v)$, where $w$ are the read-only parameters to
the first function and $v$ the non-control parameters to its
\texttt{begin} label.  Repeat execution steps until the state becomes
$\texttt{stop}(w,\,v)$, where $v$ are the non-control parameters to
the \texttt{end} label.  To run backwards, start with
$\texttt{stop}(w,\,v)$ and continue until $\texttt{start}(w,\,v)$.

Figs.~\ref{bidirectional2} and~\ref{bidirectional3} shows the rules
for function calls and returns.  Since \texttt{uncall} changes
direction of execution, there are four rules for calls and four rules
for returns.  $\mathit{CallF}$ shows a \texttt{call} in forwards
direction.  We first calculate the values $v$ and $w$ of the
eliminated and read-only arguments to the call, then find read-only
arguments and the body of the called function, then build the initial
environment from the values of the arguments.  We add the current
program point $p$ to the call stack $\sigma$ and start executing the
\texttt{Begin} block of the called function.  $\mathit{CallB}$ is the
symmetric rule for backwards execution of a \texttt{call}.

Next comes the rules for \texttt{uncall}.  Since the execution
direction is reversed, $\mathit{UncallF}$ is similar to
$\mathit{CallB}$ and $\mathit{UncallB}$ is similar to
$\mathit{CallF}$.

When returning from a function call, the direction of execution is
unchanged when returning from a \texttt{call} instruction and reversed
when returning to an \texttt{uncall} instruction. Returns to
\texttt{call} are handled in rules $\mathit{ReturnCallF}$ and
$\mathit{ReturnCallB}$, while $\mathit{ReturnUncallF}$ and
$\mathit{ReturnUncallB}$ handle returns to \texttt{uncall}.

We have tested the semantic rules by coding them in Prolog and
executed some test programs, both forwards and backwards.

\section{Contributions and Related Work}

We have defined, both informally and formally, a reversible and linear
intermediate language L4.  L4 is strongly inspired by RSSA and HSSA,
but unlike these, L4 is strongly typed and it has formal definitions
of both the type system and the execution semantics.  In its original
form~~\cite{Gail+Meyer:2024}, HSSA, like RSSA, had traditional two-way
conditional jumps, but later versions~\cite{Gail++:2026} has multi-way
control through control variables like L4.  In HSSA, every variable
that is created in a block must also be eliminated in the same block.
L4 allows variable liveness to cross block boundaries, as long as
every variable has a single creation point and a single elimination
point, and that all read-only uses of the variable happen in paths
between these points.  This makes the validity checking of L4 more
complicated than for HSSA, but it makes parameter lists shorter and
inlining of function calls easier, as you don't have to add extra
parameters to all inlined blocks.

The execution semantics is a bidirectional small-step semantics: It
contains integrated rules for forwards and backwards execution and
switching between these. The most closely related research to this is
the bidirectional small-step semantics for
Janus~\cite{Lanese+Vidal:2026}, but their approach to bidirectionality
is different: Instead of reversing execution direction at
\texttt{uncall}, they require the inverted code for the called
function to be available, and then just execute that forwards.  The
bidirectionality comes from providing both rules for (pure) forwards
execution and (pure) backward execution.  While this approach is
simpler than our approach, it is also further removed from the idea of
running the same code both forwards and backwards.  Our approach is
similar to the control mechanism in BobISA~\cite{conf/rc/ThomsenAG11}
and PISA~\cite{Frank:97}, where a direction bit determines the order
of execution and where this bit can be modified at runtime.

In addition to coding the semantic rules in Prolog, We have
implemented a parser, type-checker, and interpreter for L4 in Standard
ML.  This is available on request from the author.  The primitive
types and operations are defined in a separate module, which makes it
easy to instantiate L4 to different instruction sets.  We,
additionally, have implemented a partial evaluator for
L4~\cite{Mogensen:2027b} and a translator from a subset of L4 to the
reversible instruction set BobISA~\cite{conf/rc/ThomsenAG11} (also in
ML and also available on request).

\bibliographystyle{plain}

\bibliography{inverse,book}

@inproceedings{Alpern:1988,
author = {Alpern, B. and Wegman, M. N. and Zadeck, F. K.},
title = {Detecting equality of variables in programs},
year = {1988},
isbn = {0897912527},
publisher = {Association for Computing Machinery},
address = {New York, NY, USA},
url = {https://doi.org/10.1145/73560.73561},
doi = {10.1145/73560.73561},
booktitle = {Proceedings of the 15th ACM SIGPLAN-SIGACT Symposium on Principles of Programming Languages},
pages = {1–11},
numpages = {11},
location = {San Diego, California, USA},
series = {POPL '88}
}

@inproceedings{Rosen:1988,
author = {Rosen, B. K. and Wegman, M. N. and Zadeck, F. K.},
title = {Global value numbers and redundant computations},
year = {1988},
isbn = {0897912527},
publisher = {Association for Computing Machinery},
address = {New York, NY, USA},
url = {https://doi.org/10.1145/73560.73562},
doi = {10.1145/73560.73562},
booktitle = {Proceedings of the 15th ACM SIGPLAN-SIGACT Symposium on Principles of Programming Languages},
pages = {12–27},
numpages = {16},
location = {San Diego, California, USA},
series = {POPL '88}
}

@inproceedings{Gail+Meyer:2024,
  author       = {Lukas Gail and
                  Uwe Meyer},
  editor       = {Torben {\AE}gidius Mogensen and
                  Lukasz Mikulski},
  title        = {Connecting Reversible and Classical Computing Through Hybrid {SSA}},
  booktitle    = {Reversible Computation - 16th International Conference, {RC} 2024,
                  Toru{\'{n}}, Poland, July 4-5, 2024, Proceedings},
  series       = {Lecture Notes in Computer Science},
  volume       = {14680},
  pages        = {161--178},
  publisher    = {Springer},
  year         = {2024},
  url          = {https://doi.org/10.1007/978-3-031-62076-8\_11},
  doi          = {10.1007/978-3-031-62076-8\_11},
  bibsource    = {dblp computer science bibliography, https://dblp.org}
}

@inproceedings{matsakis:2014,
  title={The {Rust} language},
  author={Matsakis, Nicholas D and Klock II, Felix S},
  booktitle={ACM SIGAda Ada Letters},
  volume={34},
  pages={103--104},
  year={2014},
  organization={ACM}
}

@inproceedings{conf/rc/ThomsenAG11,
  author = {Thomsen, Michael Kirkedal and Axelsen, Holger Bock and Glück, Robert},
  booktitle = {RC 2011},
  editor = {Vos, Alexis De and Wille, Robert},
  ee = {http://dx.doi.org/10.1007/978-3-642-29517-1_3},
  isbn = {978-3-642-29516-4},
  pages = {30-42},
  publisher = {Springer},
  series = {Lecture Notes in Computer Science},
  title = {A Reversible Processor Architecture and Its Reversible Logic Design.},
  url = {http://dblp.uni-trier.de/db/conf/rc/rc2011.html#ThomsenAG11},
  volume = 7165,
  year = 2011
}

@incollection{Mogensen:RC2015,
year={2015},
booktitle={Proceedings of Reversible Computation 2015 (RC 2015)},
series={Lecture Notes in Computer Science},
volume = {9138},
pages = {79-94},
title={Garbage Collection for Reversible Functional Languages},
publisher={Springer Berlin Heidelberg},
author={Mogensen, Torben {\AE}gidius},
}

@incollection{Mogensen:PSI2015,
year={2016},
booktitle={Perspectives of System Informatics - 10th International Ershov Informatics Conference, {PSI} 2015, Kazan, Russia, August 24-27, 2015},
title={{RSSA}: A Reversible {SSA} Form},
series={Lecture Notes in Computer Science},
volume = {9609},
pages = {203-217},
publisher={Springer Berlin Heidelberg},
author={Mogensen, Torben {\AE}gidius},
}

@inproceedings{Mogensen:2021,
title = "Reversible Functional Array Programming",
author = "Mogensen, {Torben {\AE}gidius}",
year = "2021",
doi = "10.1007/978-3-030-79837-6_3",
language = "English",
isbn = "9783030798369",
series = "Lecture Notes in Computer Science",
publisher = "Springer",
pages = "45--63",
editor = "Shigeru Yamashita and Tetsuo Yokoyama",
booktitle = "Reversible Computation 13th International - 13th International Conference, RC 2021, Proceedings",
address = "Switzerland",
}

@InProceedings{Lanese+Vidal:2026,
author="Lanese, Ivan
and Vidal, Germ{\'a}n",
editor="Aubert, Cl{\'e}ment
and Roversi, Luca",
title="A Reversible Semantics for {Janus}",
booktitle={Reversible Computation: 18th International Conference, RC 2026, Turin, Italy, July 9–10, 2026, Proceedings},
year="2026",
publisher="Springer Nature Switzerland",
address="Cham",
pages="184--200",
isbn="978-3-032-30839-9"
}

@Misc{Frank:97,
  AUTHOR       = {Michael P. Frank},
  title =        {Modifications to {PISA} architecture to support guaranteed reversibility and other features},
  YEAR         = {1997},
  howpublished =  {MIT Reversible Computing Project Memo \#M7},
  url = {http://revcomp.info/legacy/mpf/rc/memos/M07/M07_revarch.html},
}

@inproceedings{Gail++:2026,
author = {Gail, Lukas and Meyer, Uwe and Sch\"{o}nhals, Tristan},
title = {Compiling {Roopl++} to {HSSA}},
year = {2026},
editor="Aubert, Cl{\'e}ment
and Roversi, Luca",
isbn = {978-3-032-30838-2},
publisher = {Springer-Verlag},
address = {Berlin, Heidelberg},
url = {https://doi.org/10.1007/978-3-032-30839-9_10},
doi = {10.1007/978-3-032-30839-9_10},
booktitle = {Reversible Computation: 18th International Conference, RC 2026, Turin, Italy, July 9–10, 2026, Proceedings},
pages = {167–183},
numpages = {17},
location = {Turin, Italy}
}

@Misc{Mogensen:2027b,
  AUTHOR       = {Torben Æ. Mogensen},
  title =        {Partial Evaluation of the Reversible Intermediate
  Language {L4}},
  YEAR         = {2027},
  howpublished =  {To appear},
}

\end{document}